\documentclass[aps,superscriptaddress,reprint,longbibliography]{revtex4-1}
\usepackage{graphicx}
\begin{document}

\title{Student Responses to Changes in Introductory Physics Learning due to COVID-19 Pandemic}

\date{\today}

\author{Matthew Dew}
\affiliation{Department of Physics and Astronomy, Texas A\&M University,
		College Station, TX~~77843}
\email[send correspondence to: ]{etanya@tamu.edu}
\author{Jonathan Perry}
\affiliation{Department of Physics, University of Texas,
		Austin, TX~~78712}
\author{Lewis Ford}
\affiliation{Department of Physics and Astronomy, Texas A\&M University,
		College Station, TX~~77843}
\author{Dawson Nodurft}
\affiliation{Department of Physics and Astronomy, Texas A\&M University,
		College Station, TX~~77843}
		\author{Tatiana Erukhimova}
\affiliation{Department of Physics and Astronomy, Texas A\&M University,
		College Station, TX~~77843}

\maketitle

As a result of the spread of COVID-19 during spring 2020, many colleges and universities across the US, and beyond, were compelled to move entirely to remote, online instruction, or shut down \cite{unesco}. Due to the rapidity of this transition, instructors had to significantly -- if not completely -- change their instructional style on very short notice \cite{Hodges2020}. Our purpose with this paper is to report on student experiences and reactions to the switch to emergency remote learning at two large, land-grant, research intensive universities. We aimed to explore how students have received and dealt with the shift to remote learning that began in March 2020, specifically in introductory physics and astronomy courses. By providing timely student feedback, we hope to help instructors tune their efforts to build a more effective remote learning environment.

While online courses have been around for several decades, most major universities do not offer more than a handful of their classes online \cite{NationalCenterForEducationStats}. Existing literature on the principles of design and evaluation of online courses suggest that high-quality online courses require several months of preparation as well as infrastructure and support for both students and instructors \cite{McGee2017, Hodges2020}. Even with the careful preparation of an online course, studies prior to the 2020 pandemic have often revealed a deficiency of collaborative aspects, in both student-student and student-instructor interactions \cite{Paulsen2020, Lowenthal2017, GovReport}. Literature focused on online introductory physics, or courses with online components, show these interactions being linked to student success \cite{Faulconer2018, Miller2016}. While the educational community put in extraordinary effort in spring 2020 to keep teaching, the timeline of the shift to emergency remote learning did not allow for the thoughtful planning of a typical online course.

Given how fast and total the switch to remote learning has been, we do not fully understand how students have been affected -- both in general and for specific student populations. This transition can exacerbate inequity and disadvantage students from lower income families \cite{PEW2019}. Pandemic-induced isolation and socio-economic hardships may also affect students’ mental health \cite{Chen2020}. 

To explore how students responded to the shift to emergency remote learning, we developed and administered a questionnaire gauging the impacts on students motivation and interactions with their courses, their peers, and instructors. We also examined how student responses depend on demographic factors which are listed below. 

\begin{center}
    \textbf{METHODS}
\end{center}

The questionnaire was distributed to 2,320 students enrolled in summer 2020 courses, which were delivered remotely, at both participating institutions. At Institution A, the questionnaire was distributed via email by departmental administration. Additionally, a member of the research team attended a lecture for each class encouraging students to take the questionnaire. Here, there were a total of 13 lecture sections from 6 courses taught by 8 instructors -- covering both algebra-based and calculus-based introductory physics, an introductory astronomy course, and a sophomore-level modern physics course. At Institution B, the questionnaire was distributed by faculty via the Learning Management System. Here, there were 6 lecture sections from 6 courses taught by 5 instructors -- covering three, two semester tracks of algebra-based and calculus-based introductory physics. 

The questionnaire was entirely anonymous. Student participation was voluntary with no positive or negative inducements. Distribution occurred during the second half of courses, which were primarily 5-week courses (one course was 10 weeks). A total of 708 responses were received, 508 from Institution A (37.5\% response rate), and 200 from Institution B (20.7\% response rate).

The questionnaire consisted of 8 demographics questions and 27 questions on student preferences and behaviors. The 8 demographics questions asked for student gender, race, first-generation status, classification, weekly work hours, internet access, and devices used for classwork. Questionnaire items were built with three different response types: 5-point Likert scale, multiple answer, and ranking. 

Each of the response types mentioned above was examined in a distinct manner. Likert-scale items are represented by the fraction of students who chose each answer, summing to 1. Multiple answer items are represented by the fraction of responses for each individual choice. Ranking items assigned a point value to each answer, where each response's score was normalized by the total number of points distributed. The 95\% confidence intervals were calculated and are included on each figure.

\pagebreak
\begin{center}
    \textbf{RESULTS \& DISCUSSION}
\end{center}

\subparagraph{Demographics:} Students responding to the questionnaire were more likely to be female (59.8\%) than male (38.9\%). The remaining percentage of respondents identified as non-binary or preferred not to say. Nearly three-quarters of responses came from juniors and sophomores. Students primarily identified as white (58.6\%), with Asian and Hispanic/Latinx/Spanish origin identifying students responding in roughly equal numbers (23.7\% and 22.7\% respectively). Students were allowed to select more than one race or ethnicity and were counted in all categories selected. Around one in five students identified as a first-generation college student.

\begin{figure}[htp]
\centering
\includegraphics[width=8.6cm]{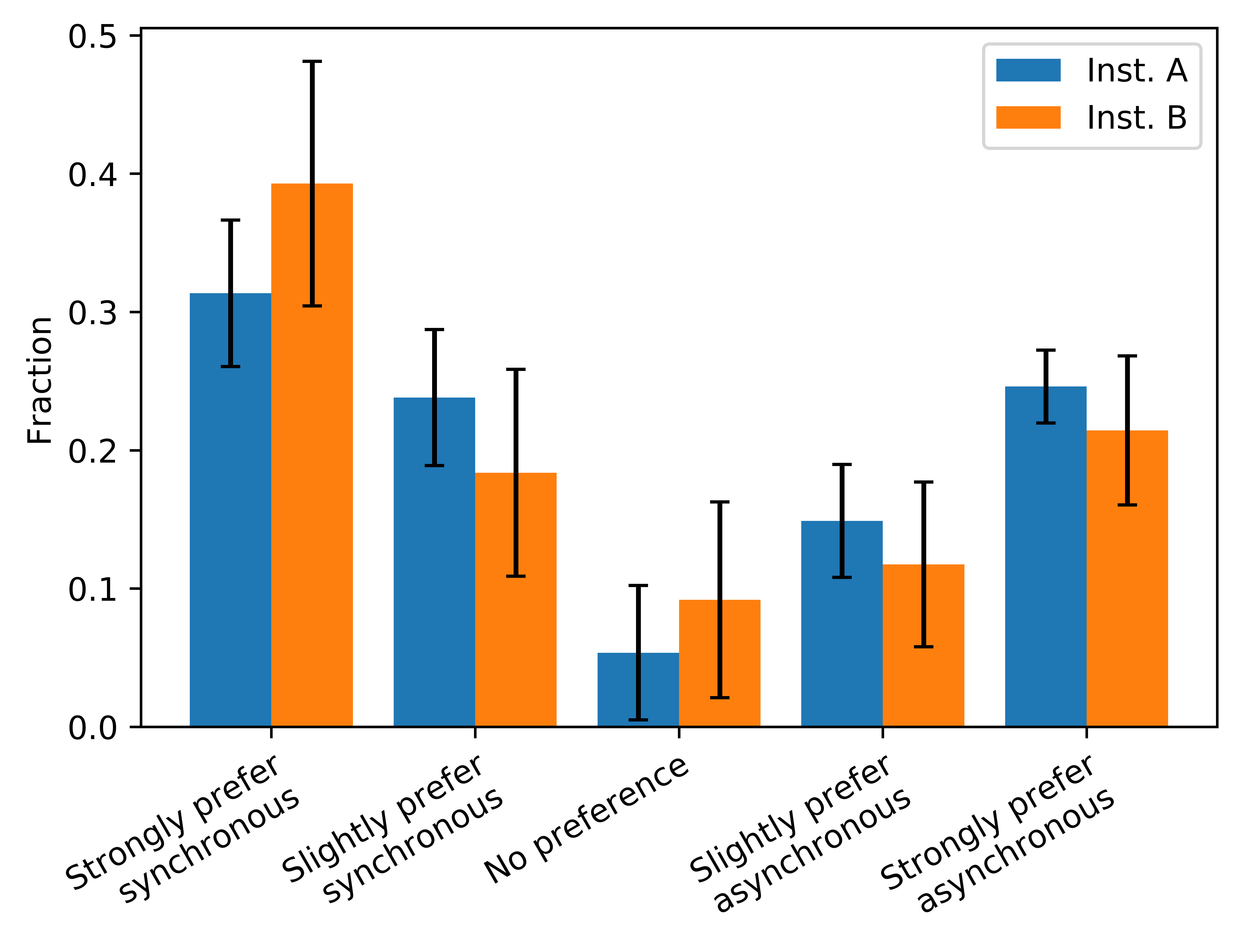}
\caption{\label{tab:SynVsAsyn}Student responses to ``Between synchronous classes and asynchronous classes, what do you prefer for lecture?'' }
\end{figure}

\begin{figure}[htp]
\centering
\includegraphics[width=8.6cm]{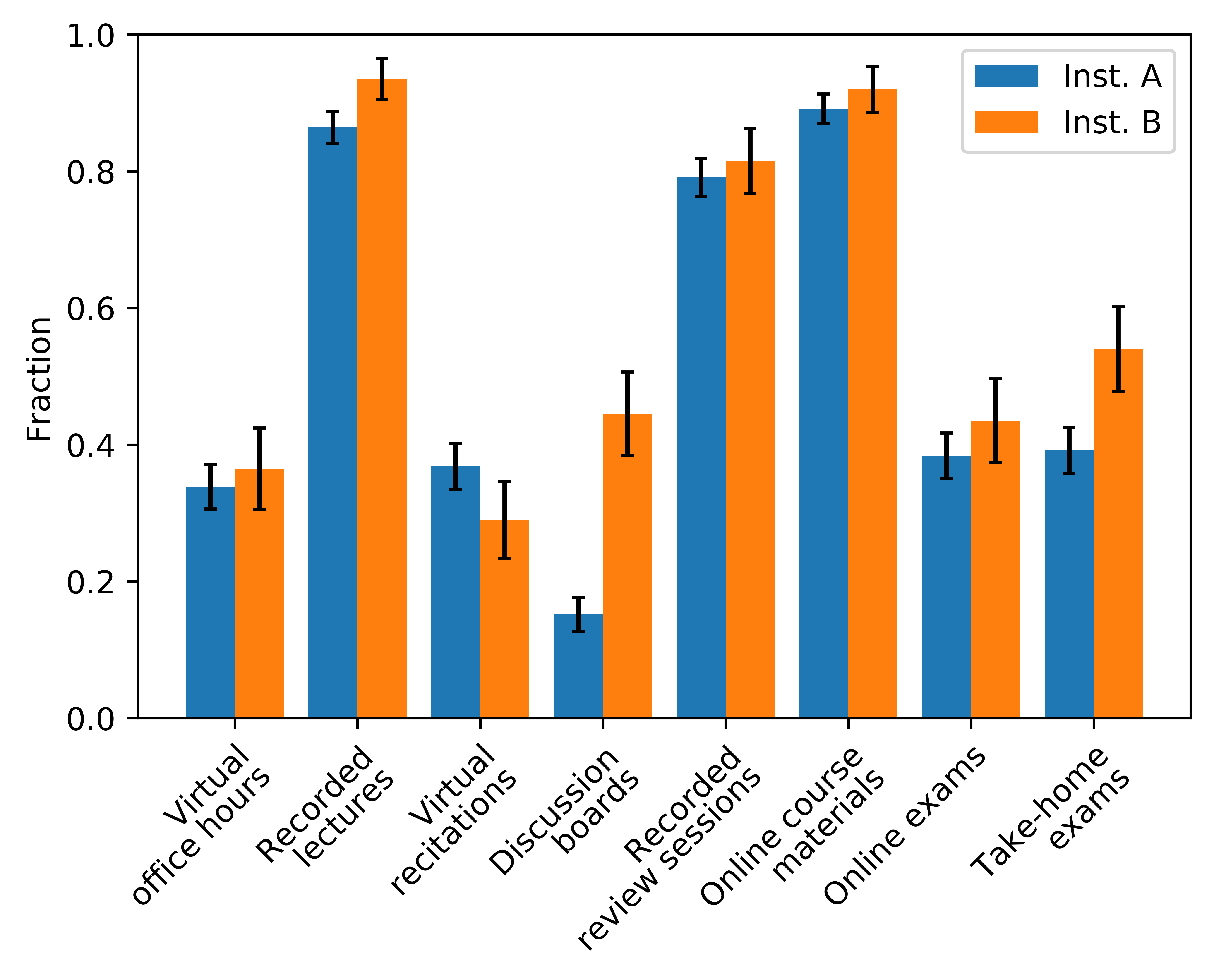}
\caption{\label{tab:Q10} ``Which part(s) of online classes do you want to keep when classes return to normal?''}
\end{figure}

\subparagraph{Synchronous vs. Asynchronous:} The majority of the opinions expressed on this item are in the “strongly prefer” categories, indicating a strong polarization among respondents. Overall, students at Institution A exhibited a slight preference for synchronous classes over asynchronous classes; more students at Institution B showed a stronger preference (around 2-to-1), Figure \ref{tab:SynVsAsyn}. Perhaps related to the preference for synchronous learning, students had a strong agreement ($\sim$70\%) when asked if they would have benefited more from face-to-face instruction. From both institutions more than 75\% of respondents indicated that they would like to continue to have access to recorded lectures and review sessions, as well as online course materials when classes return to a face-to-face format, Figure \ref{tab:Q10}. It would appear then, that students see a value in live instruction and engagement from faculty while desiring the convenience of on-demand access to course information while studying.

\subparagraph{Remote Learning Capabilities:} About 40\% of students do not have their own study space to work undisturbed at least most of the time. Shared spaces may mean that these students do not have privacy to necessarily feel comfortable using a microphone or webcamera during every component of a course. Though all responding students have access to at least a desktop or laptop computer, less than half reported access to a printer, and even fewer reported access to a scanner (22.5\%) or a document camera (10.5\%). It should be noted that both institutions made specific efforts to provide technology (including purchase of laptops) for students during the spring semester.

Resources such as textbooks, recitations, and instructor-provided online resources have become more important to student learning since the pandemic. Students report that they are most likely to ask a friend, or reference online problem-solving resources (\textit{e.g.} Chegg) when stuck on a problem. This behavior is not surprising, and may not be significantly different than student behavior during a typical semester. However, it is important context when considering study habits, which are discussed below.

\begin{figure}[htp]
\centering
\includegraphics[width=8.6cm]{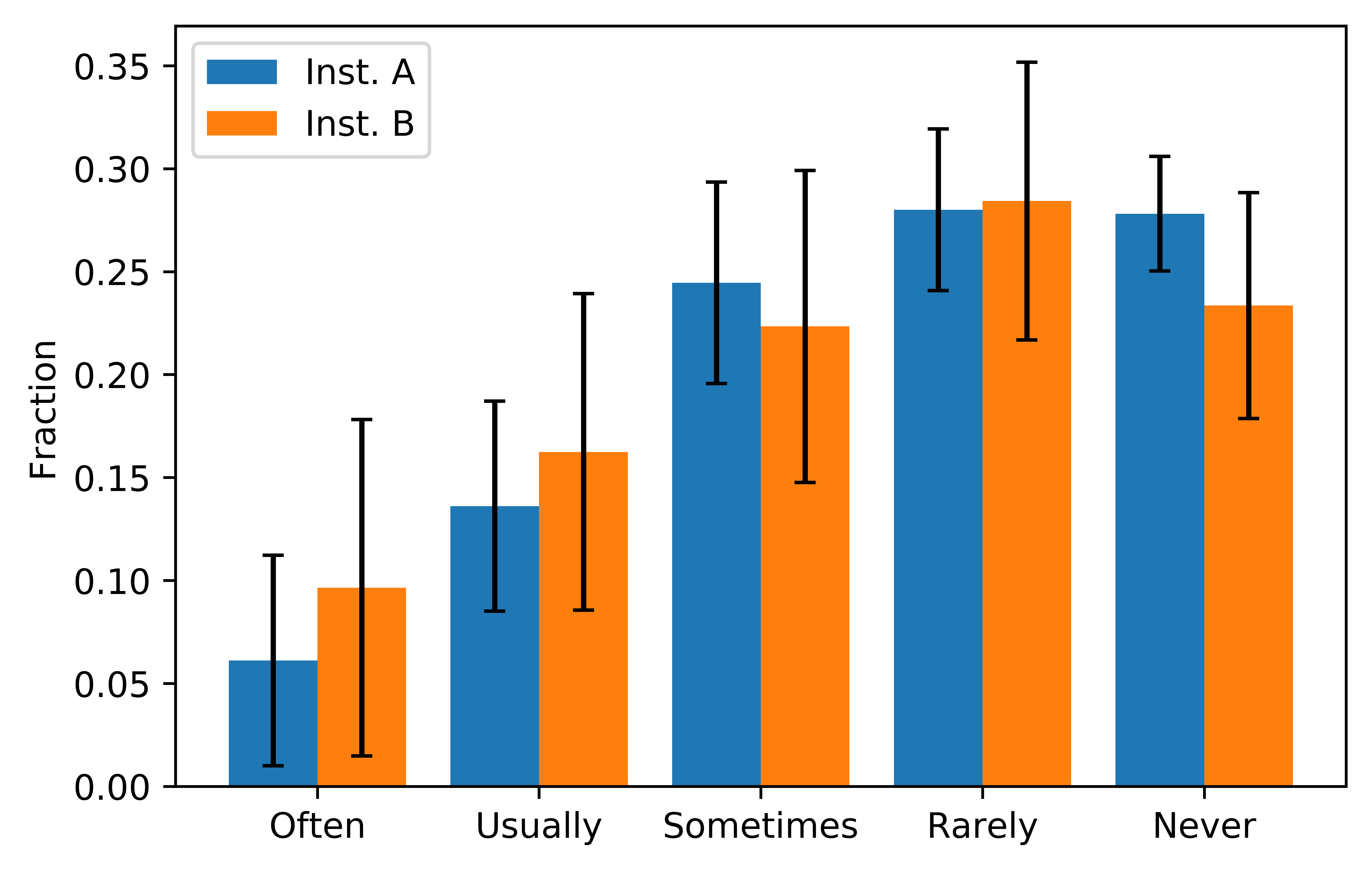}
\caption{\label{tab:Q43-11} Student responses to ``I felt engaged/connected with my classmates during this course's lecture.''}
\end{figure}

\subparagraph{Community \& Engagement:} On the whole, students did not feel connected or engaged with their classmates during their 2020 summer physics courses. Over half of the students reported feeling ``rarely'' or ``never'' connected or engaged with classmates during lecture, Figure \ref{tab:Q43-11}. These responses are similar for all components of the course that students were asked about. This likely strongly contributed to more than 50\% of students reporting far fewer social connections being made in this course compared to their average face-to-face classes. 

\begin{figure}[htp]
\centering
\includegraphics[width=8.6cm]{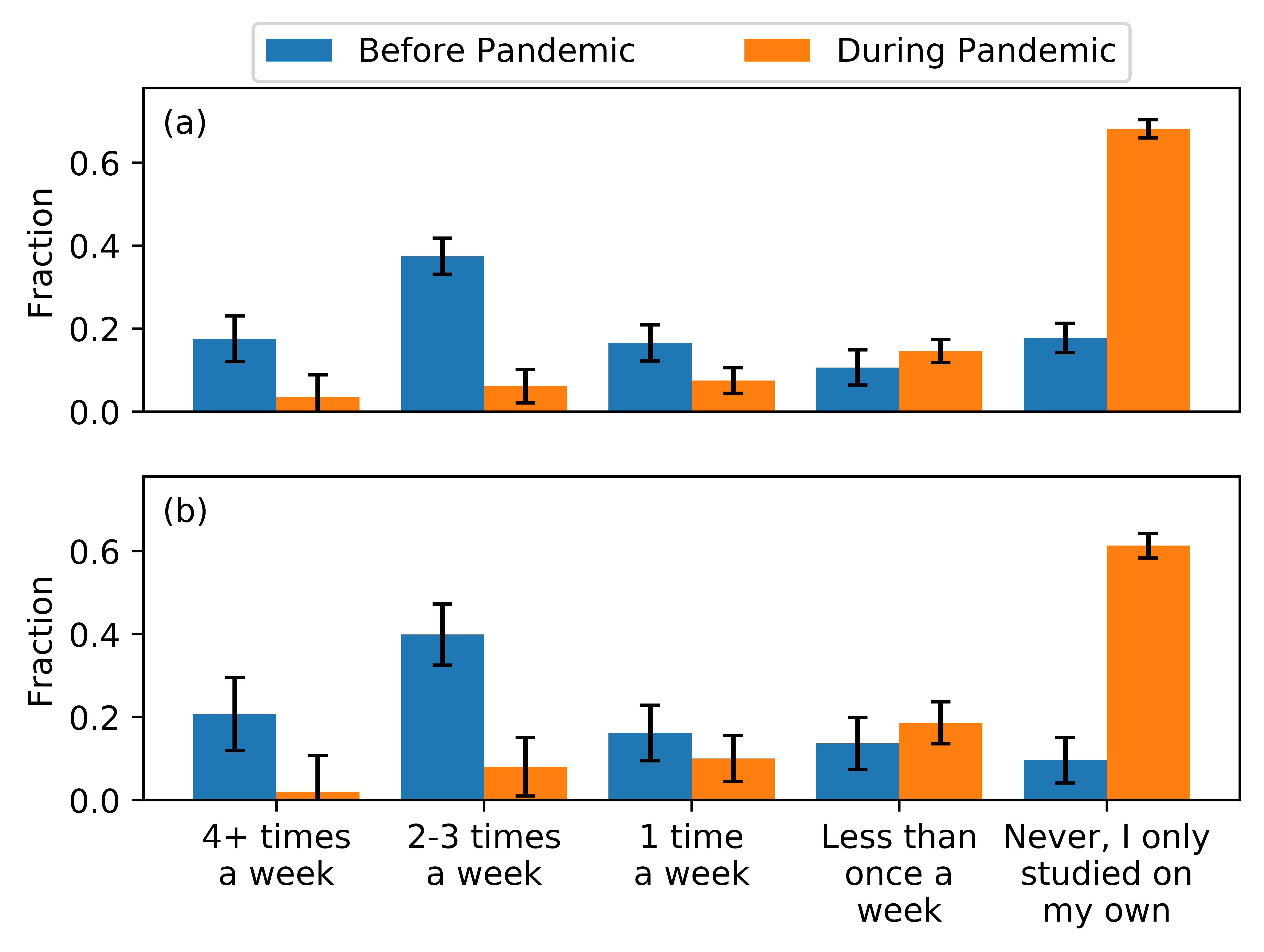}
\caption{\label{tab:Q27andQ28} Student responses to ``On average, for each class, how many times a week did you meet with classmates to study before/during the pandemic?'' for (a) Institution A and (b) Institution B.}
\end{figure}

Student study habits changed drastically in the social components of their courses. Before the pandemic, about two thirds of the students met at least once a week to study. During the pandemic, however, nearly the same fraction of students reported studying only alone, Figure \ref{tab:Q27andQ28}. While virtual meetings allow students to communicate and potentially study together, the ability to collaborate and share work in real time has become more difficult.

Despite the decrease in group studying, one fourth of students never contacted their instructor outside of class. Many students who attended office hours multiple times during the course were neutral on the visits’ usefulness. About 35\% found them helpful.

One method of engaging students during lecture, in-class demonstrations, received significant agreement ($\sim$60\%) for aiding in student understanding or maintaining interest in class. Responses to these items are muddled as about 20\% of students from each course stated that no demonstrations were used. One pre-pandemic study suggests that video versions of physics demonstrations can be an effective contribution to student learning \cite{kestin2020comparing}. 

\begin{figure}[htp]
\centering
\includegraphics[width=8.6cm]{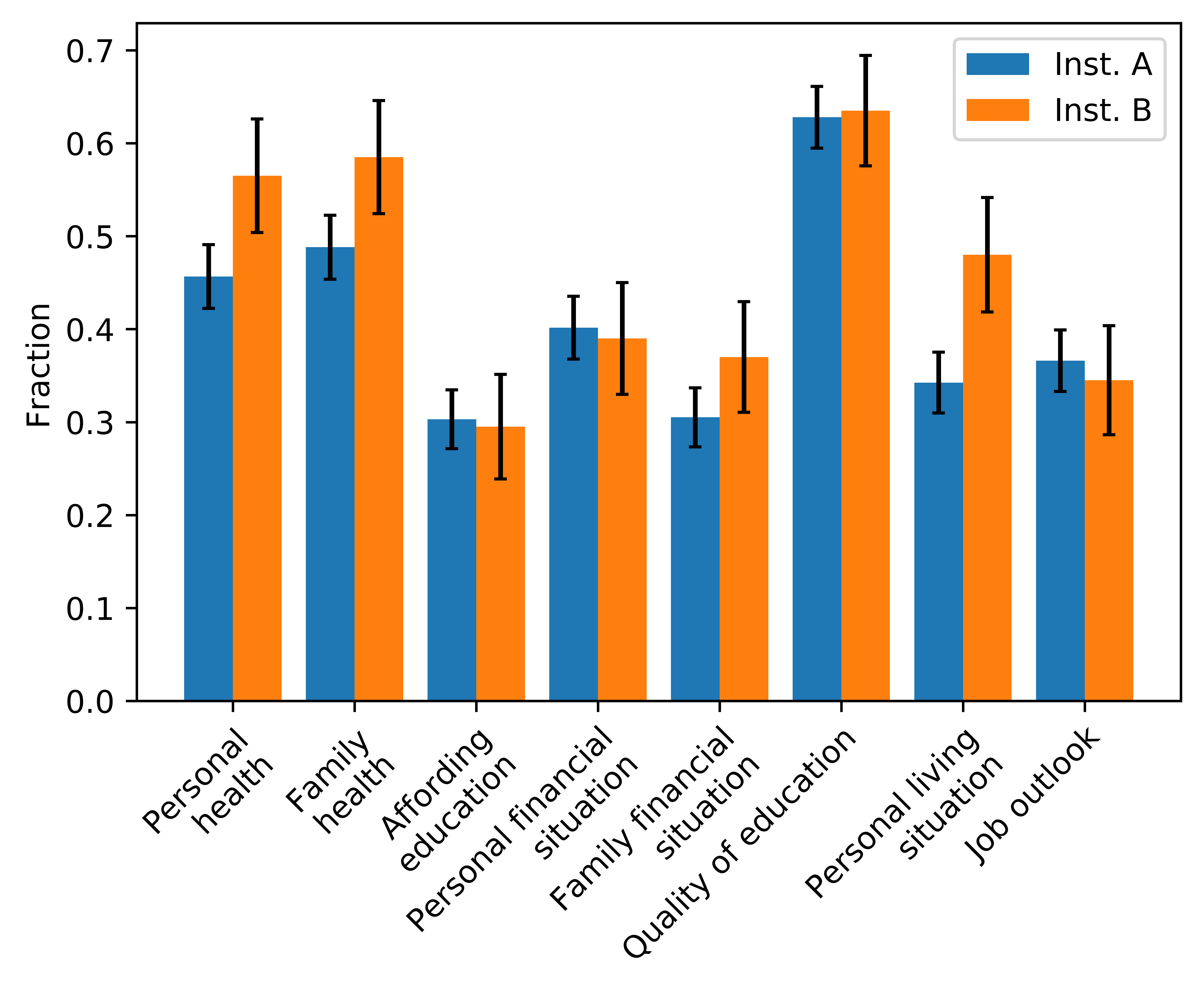}
\caption{\label{tab:Q39} Student responses to ``Which of these issues have been a cause for stress for you since the pandemic? Mark all that are applicable.'' This is plotted by the fraction of students reporting this cause of stress.}
\end{figure}

\subparagraph{Stress \& Motivation:} Students reported quality of education as the most common cause of stress at both institutions (63\%), Figure \ref{tab:Q39}. This is followed by an expected feeling of stress about personal and family health, both around 50\%. It is surprising that in a time of pandemic more students reported stress relating to their education than their own health.

\begin{figure}[htp]
\centering
\includegraphics[width=8.6cm]{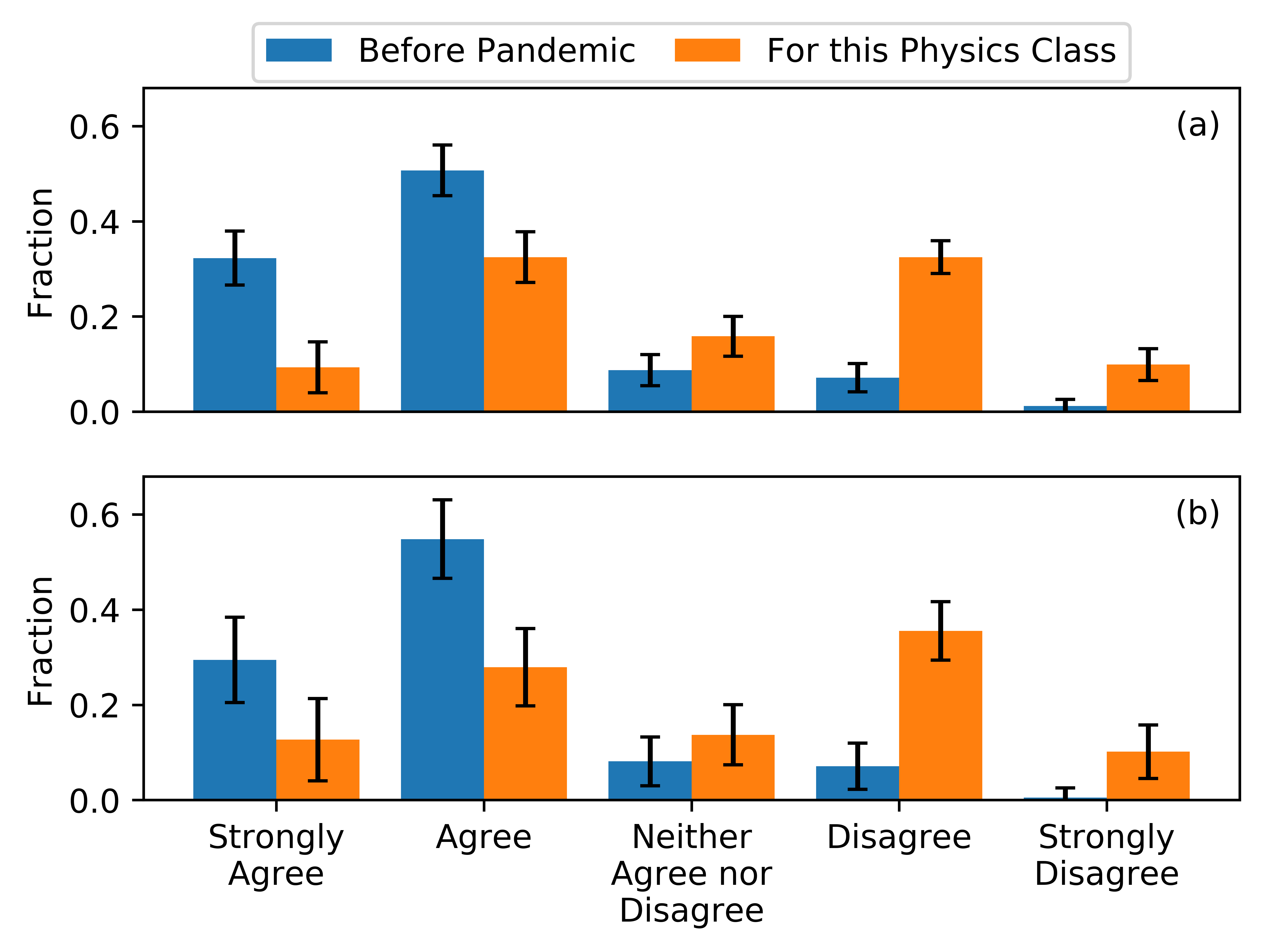}
\caption{\label{tab:Q43-21 and Q43-22} Student responses to ``Before the pandemic/For this physics class, I was able to get myself to work on assignments without difficulty.'' for (a) Institution A and (b) Institution B.}
\end{figure}

Perhaps related to student stress over education, reported motivation for being able to work on assignments changed drastically after shifting to remote learning. Figure \ref{tab:Q43-21 and Q43-22} shows a striking decline in the ease of students being able to work on their course assignments from both institutions. One should keep in mind this is not a direct comparison between two physics courses taken face-to-face and remotely. The ``before pandemic'' question includes students who have never been enrolled in a physics course.

Both Institution A and Institution B experienced a significant increase in enrollment during summer courses, more than doubling from prior years. While many responding students stated they would have taken the courses at these institutions regardless of circumstances (48\%), other students changed their plans; either not enrolling at an alternative institution (19\%), or taking the course because it was offered online (18\%).

\subparagraph{Differences by Demographics:} We examined responses to each questionnaire item separated by student demographics based on gender, first-generation status, classification, employment, course enrolled in, and primary motivation. Only one identifier, first-generation status, led to noteworthy differences from what has been mentioned earlier.

\begin{figure}[htp]
\centering
\includegraphics[width=8.6cm]{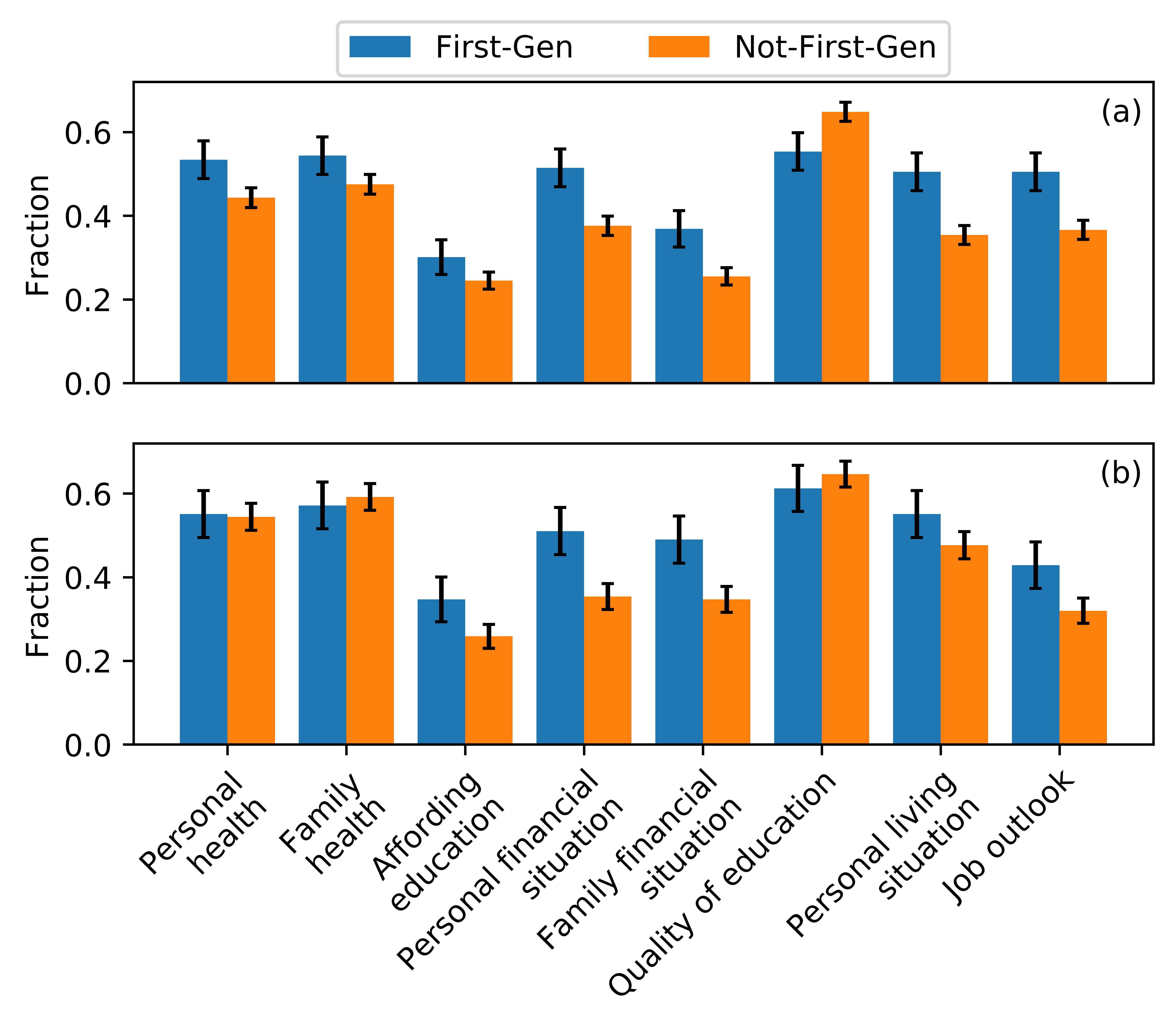}
\caption{\label{tab:Q39ConditionFromQ4} Student responses to ``Which of these issues have been a cause for stress for you since the pandemic?'' for (a) Institution A and (b) Institution B. These responses are separated by student responses to ``Are you a first generation college student?''}
\end{figure}

First-generation students reported more frequently than non-first-generation students every listed cause of stress except for quality of education, Figure \ref{tab:Q39ConditionFromQ4}. First-generation students are also less likely to have access to a private study space, 52\% compared to 74\% of non-first generation students.

\subparagraph{Summary:} In response to the sudden and singular changes during spring 2020 necessitated by the pandemic, we developed and administered a questionnaire with the goal of finding ways to better serve our students. We collected student feedback from summer 2020 introductory physics and astronomy courses at two large, land-grant institutions. Analysis of responses shows that students are experiencing drastically reduced social interactions and connections through their courses. Individual motivation to complete coursework has decreased precipitously. Students are also experiencing multiple causes of stress, with quality of education topping the list. These reactions are expressed across all demographic categories in a similar way. Noticeable differences are observed only for first-generation students who report causes of stress more often and are less likely to have a private study space. Post-pandemic, increased availability of course resources, particularly recorded lectures and online materials, resonates with students.

Although a direct comparison cannot be made between emergency remote learning and online education, student responses about interactions within courses mirror pre-pandemic literature studying online learning \cite{Offir2008, Paulsen2020, Miller2016}. Readers should keep in mind that responses to the questionnaire represent only a snapshot of student preferences and experiences gathered during condensed versions of physics and astronomy courses during summer 2020. In reporting this, we have aimed to provide a slice of timely information as educators continue to make instructional choices during remote learning and eventually, for the return to face-to-face classes.

\begin{acknowledgments}
We would like to thank the faculty from both institutions who allowed us to distribute the questionnaire to their students during their summer courses. This study was supported in part by the Texas A\&M University Department of Physics \& Astronomy.
\end{acknowledgments}

\bibliography{bibliography}

\begin{thebibliography}{13}%
\makeatletter
\providecommand \@ifxundefined [1]{%
 \@ifx{#1\undefined}
}%
\providecommand \@ifnum [1]{%
 \ifnum #1\expandafter \@firstoftwo
 \else \expandafter \@secondoftwo
 \fi
}%
\providecommand \@ifx [1]{%
 \ifx #1\expandafter \@firstoftwo
 \else \expandafter \@secondoftwo
 \fi
}%
\providecommand \natexlab [1]{#1}%
\providecommand \enquote  [1]{``#1''}%
\providecommand \bibnamefont  [1]{#1}%
\providecommand \bibfnamefont [1]{#1}%
\providecommand \citenamefont [1]{#1}%
\providecommand \href@noop [0]{\@secondoftwo}%
\providecommand \href [0]{\begingroup \@sanitize@url \@href}%
\providecommand \@href[1]{\@@startlink{#1}\@@href}%
\providecommand \@@href[1]{\endgroup#1\@@endlink}%
\providecommand \@sanitize@url [0]{\catcode `\\12\catcode `\$12\catcode
  `\&12\catcode `\#12\catcode `\^12\catcode `\_12\catcode `\%12\relax}%
\providecommand \@@startlink[1]{}%
\providecommand \@@endlink[0]{}%
\providecommand \url  [0]{\begingroup\@sanitize@url \@url }%
\providecommand \@url [1]{\endgroup\@href {#1}{\urlprefix }}%
\providecommand \urlprefix  [0]{URL }%
\providecommand \Eprint [0]{\href }%
\providecommand \doibase [0]{http://dx.doi.org/}%
\providecommand \selectlanguage [0]{\@gobble}%
\providecommand \bibinfo  [0]{\@secondoftwo}%
\providecommand \bibfield  [0]{\@secondoftwo}%
\providecommand \translation [1]{[#1]}%
\providecommand \BibitemOpen [0]{}%
\providecommand \bibitemStop [0]{}%
\providecommand \bibitemNoStop [0]{.\EOS\space}%
\providecommand \EOS [0]{\spacefactor3000\relax}%
\providecommand \BibitemShut  [1]{\csname bibitem#1\endcsname}%
\let\auto@bib@innerbib\@empty
\bibitem [{\citenamefont {UNESCO}(2020)}]{unesco}%
  \BibitemOpen
  \bibfield  {author} {\bibinfo {author} {\bibnamefont {UNESCO}},\ }\bibfield
  {title} {\enquote {\bibinfo {title} {Covid-19 educational disruption and
  response},}\ }\href@noop {} {\  (\bibinfo {year} {2020})},\ \bibinfo {note}
  {retrieved from https://en.unesco.org/covid19/educationresponse}\BibitemShut
  {NoStop}%
\bibitem [{\citenamefont {Hodges}\ \emph {et~al.}(2020)\citenamefont {Hodges},
  \citenamefont {Moore}, \citenamefont {Lockee}, \citenamefont {Trust},\ and\
  \citenamefont {Bond}}]{Hodges2020}%
  \BibitemOpen
  \bibfield  {author} {\bibinfo {author} {\bibfnamefont {Charles}\ \bibnamefont
  {Hodges}}, \bibinfo {author} {\bibfnamefont {Stephanie}\ \bibnamefont
  {Moore}}, \bibinfo {author} {\bibfnamefont {Barb}\ \bibnamefont {Lockee}},
  \bibinfo {author} {\bibfnamefont {Torrey}\ \bibnamefont {Trust}}, \ and\
  \bibinfo {author} {\bibfnamefont {Aaron}\ \bibnamefont {Bond}},\ }\href
  {https://er.educause.edu/articles/2020/3/the-difference-between-emergency-remote-teaching-and-online-learning}
  {} (\bibinfo {year} {2020}),\ \bibinfo {note} {retrieved from
  https://er.educause.edu/articles/2020/3/the-difference-between-emergency-remote-teaching-and-online-learning}\BibitemShut
  {NoStop}%
\bibitem [{\citenamefont {for
  Education~Statistics}(2019)}]{NationalCenterForEducationStats}%
  \BibitemOpen
  \bibfield  {author} {\bibinfo {author} {\bibfnamefont {National~Center}\
  \bibnamefont {for Education~Statistics}},\ }\href@noop {} {\enquote {\bibinfo
  {title} {Number and percentage of students enrolled in degree-granting
  postsecondary institutions, by distance education participation, location of
  student, level of enrollment, and control and level of institution: Fall 2017
  and fall 2018},}\ } (\bibinfo {year} {2019})\BibitemShut {NoStop}%
\bibitem [{\citenamefont {McGee}\ \emph {et~al.}(2017)\citenamefont {McGee},
  \citenamefont {Windes},\ and\ \citenamefont {Torres}}]{McGee2017}%
  \BibitemOpen
  \bibfield  {author} {\bibinfo {author} {\bibfnamefont {Patricia}\
  \bibnamefont {McGee}}, \bibinfo {author} {\bibfnamefont {Deborah}\
  \bibnamefont {Windes}}, \ and\ \bibinfo {author} {\bibfnamefont {Maria}\
  \bibnamefont {Torres}},\ }\bibfield  {title} {\enquote {\bibinfo {title}
  {Experienced online instructors: beliefs and preferred supports regarding
  online teaching},}\ }\href {\doibase 10.1007/s12528-017-9140-6} {\bibfield
  {journal} {\bibinfo  {journal} {Journal of Computing in Higher Education}\
  }\textbf {\bibinfo {volume} {29}},\ \bibinfo {pages} {331--352} (\bibinfo
  {year} {2017})}\BibitemShut {NoStop}%
\bibitem [{\citenamefont {Paulsen}\ and\ \citenamefont
  {McCormick}(2020)}]{Paulsen2020}%
  \BibitemOpen
  \bibfield  {author} {\bibinfo {author} {\bibfnamefont {Justin}\ \bibnamefont
  {Paulsen}}\ and\ \bibinfo {author} {\bibfnamefont {Alexander~C.}\
  \bibnamefont {McCormick}},\ }\bibfield  {title} {\enquote {\bibinfo {title}
  {Reassessing disparities in online learner student engagement in higher
  education},}\ }\href {\doibase 10.3102/0013189X19898690} {\bibfield
  {journal} {\bibinfo  {journal} {Educational Researcher}\ }\textbf {\bibinfo
  {volume} {49}},\ \bibinfo {pages} {20--29} (\bibinfo {year}
  {2020})}\BibitemShut {NoStop}%
\bibitem [{\citenamefont {Lowenthal}\ and\ \citenamefont
  {Snelson}(2017)}]{Lowenthal2017}%
  \BibitemOpen
  \bibfield  {author} {\bibinfo {author} {\bibfnamefont {Patrick}\ \bibnamefont
  {Lowenthal}}\ and\ \bibinfo {author} {\bibfnamefont {Chareen}\ \bibnamefont
  {Snelson}},\ }\bibfield  {title} {\enquote {\bibinfo {title} {In search of a
  better understanding of social presence: an investigation into how
  researchers define social presence},}\ }\href {\doibase
  10.1080/01587919.2017.1324727} {\bibfield  {journal} {\bibinfo  {journal}
  {Distance Education}\ }\textbf {\bibinfo {volume} {38}},\ \bibinfo {pages}
  {1--19} (\bibinfo {year} {2017})}\BibitemShut {NoStop}%
\bibitem [{\citenamefont {Means}\ \emph {et~al.}(2010)\citenamefont {Means},
  \citenamefont {Toyama}, \citenamefont {Murphy}, \citenamefont {Bakia},
  \citenamefont {Jones},\ and\ \citenamefont {Planning}}]{GovReport}%
  \BibitemOpen
  \bibfield  {author} {\bibinfo {author} {\bibfnamefont {Barbara}\ \bibnamefont
  {Means}}, \bibinfo {author} {\bibfnamefont {Yukie}\ \bibnamefont {Toyama}},
  \bibinfo {author} {\bibfnamefont {Robert}\ \bibnamefont {Murphy}}, \bibinfo
  {author} {\bibfnamefont {Marianne}\ \bibnamefont {Bakia}}, \bibinfo {author}
  {\bibfnamefont {Karla}\ \bibnamefont {Jones}}, \ and\ \bibinfo {author}
  {\bibfnamefont {Evaluation}\ \bibnamefont {Planning}},\ }\bibfield  {title}
  {\enquote {\bibinfo {title} {Evaluation of evidence-based practices in online
  learning: A meta-analysis and review of online learning studies},}\
  }\href@noop {} {\  (\bibinfo {year} {2010})},\ \bibinfo {note} {retrieved
  from
  https://www2.ed.gov/rschstat/eval/tech/evidence-based-practices/finalreport.pdf}\BibitemShut
  {NoStop}%
\bibitem [{\citenamefont {Faulconer}\ \emph {et~al.}(2018)\citenamefont
  {Faulconer}, \citenamefont {Griffith}, \citenamefont {Wood}, \citenamefont
  {Acharyya},\ and\ \citenamefont {Roberts}}]{Faulconer2018}%
  \BibitemOpen
  \bibfield  {author} {\bibinfo {author} {\bibfnamefont {E.~K.}\ \bibnamefont
  {Faulconer}}, \bibinfo {author} {\bibfnamefont {J.}~\bibnamefont {Griffith}},
  \bibinfo {author} {\bibfnamefont {B.}~\bibnamefont {Wood}}, \bibinfo {author}
  {\bibfnamefont {S.}~\bibnamefont {Acharyya}}, \ and\ \bibinfo {author}
  {\bibfnamefont {D.}~\bibnamefont {Roberts}},\ }\bibfield  {title} {\enquote
  {\bibinfo {title} {A comparison of online, video synchronous, and traditional
  learning modes for an introductory undergraduate physics course},}\ }\href
  {\doibase 10.1007/s10956-018-9732-6} {\bibfield  {journal} {\bibinfo
  {journal} {Journal of Science Education and Technology}\ }\textbf {\bibinfo
  {volume} {27}},\ \bibinfo {pages} {404--411} (\bibinfo {year}
  {2018})}\BibitemShut {NoStop}%
\bibitem [{\citenamefont {Miller}\ \emph {et~al.}(2016)\citenamefont {Miller},
  \citenamefont {Zyto}, \citenamefont {Karger}, \citenamefont {Yoo},\ and\
  \citenamefont {Mazur}}]{Miller2016}%
  \BibitemOpen
  \bibfield  {author} {\bibinfo {author} {\bibfnamefont {Kelly}\ \bibnamefont
  {Miller}}, \bibinfo {author} {\bibfnamefont {Sacha}\ \bibnamefont {Zyto}},
  \bibinfo {author} {\bibfnamefont {David}\ \bibnamefont {Karger}}, \bibinfo
  {author} {\bibfnamefont {Junehee}\ \bibnamefont {Yoo}}, \ and\ \bibinfo
  {author} {\bibfnamefont {Eric}\ \bibnamefont {Mazur}},\ }\bibfield  {title}
  {\enquote {\bibinfo {title} {Analysis of student engagement in an online
  annotation system in the context of a flipped introductory physics class},}\
  }\href {\doibase 10.1103/PhysRevPhysEducRes.12.020143} {\bibfield  {journal}
  {\bibinfo  {journal} {Phys. Rev. Phys. Educ. Res.}\ }\textbf {\bibinfo
  {volume} {12}},\ \bibinfo {pages} {020143} (\bibinfo {year}
  {2016})}\BibitemShut {NoStop}%
\bibitem [{\citenamefont {Anderson}\ and\ \citenamefont
  {Kumar}(2019)}]{PEW2019}%
  \BibitemOpen
  \bibfield  {author} {\bibinfo {author} {\bibfnamefont {Monica}\ \bibnamefont
  {Anderson}}\ and\ \bibinfo {author} {\bibfnamefont {Madhumitha}\ \bibnamefont
  {Kumar}},\ }\href
  {https://www.pewresearch.org/fact-tank/2019/05/07/digital-divide-persists-even-as-lower-income-americans-make-gains-in-tech-adoption/}
  {\enquote {\bibinfo {title} {Digital divide persists even as lower-income
  americans make gains in tech adoption},}\ } (\bibinfo {year} {2019}),\
  \bibinfo {note} {retrieved from
  https://www.pewresearch.org/fact-tank/2019/05/07/digital-divide-persists-even-as-lower-income-americans-make-gains-in-tech-adoption/}\BibitemShut
  {NoStop}%
\bibitem [{\citenamefont {Chen}\ \emph {et~al.}(2020)\citenamefont {Chen},
  \citenamefont {Sun},\ and\ \citenamefont {Feng}}]{Chen2020}%
  \BibitemOpen
  \bibfield  {author} {\bibinfo {author} {\bibfnamefont {Bo}~\bibnamefont
  {Chen}}, \bibinfo {author} {\bibfnamefont {Jinlu}\ \bibnamefont {Sun}}, \
  and\ \bibinfo {author} {\bibfnamefont {Yi}~\bibnamefont {Feng}},\ }\bibfield
  {title} {\enquote {\bibinfo {title} {How have \uppercase{COVID}-19 isolation
  policies affected young people's mental health? evidence from chinese college
  students},}\ }\href {\doibase 10.3389/fpsyg.2020.01529} {\bibfield  {journal}
  {\bibinfo  {journal} {Frontiers in Psychology}\ }\textbf {\bibinfo {volume}
  {11}},\ \bibinfo {pages} {1529} (\bibinfo {year} {2020})}\BibitemShut
  {NoStop}%
\bibitem [{\citenamefont {Kestin}\ \emph {et~al.}(2020)\citenamefont {Kestin},
  \citenamefont {Miller}, \citenamefont {McCarty}, \citenamefont {Callaghan},\
  and\ \citenamefont {Deslauriers}}]{kestin2020comparing}%
  \BibitemOpen
  \bibfield  {author} {\bibinfo {author} {\bibfnamefont {Greg}\ \bibnamefont
  {Kestin}}, \bibinfo {author} {\bibfnamefont {Kelly}\ \bibnamefont {Miller}},
  \bibinfo {author} {\bibfnamefont {Logan~S}\ \bibnamefont {McCarty}}, \bibinfo
  {author} {\bibfnamefont {Kristina}\ \bibnamefont {Callaghan}}, \ and\
  \bibinfo {author} {\bibfnamefont {Louis}\ \bibnamefont {Deslauriers}},\
  }\bibfield  {title} {\enquote {\bibinfo {title} {Comparing the effectiveness
  of online versus live lecture demonstrations},}\ }\href@noop {} {\bibfield
  {journal} {\bibinfo  {journal} {Physical Review Physics Education Research}\
  }\textbf {\bibinfo {volume} {16}},\ \bibinfo {pages} {013101} (\bibinfo
  {year} {2020})}\BibitemShut {NoStop}%
\bibitem [{\citenamefont {Offir}\ \emph {et~al.}(2008)\citenamefont {Offir},
  \citenamefont {Lev},\ and\ \citenamefont {Bezalel}}]{Offir2008}%
  \BibitemOpen
  \bibfield  {author} {\bibinfo {author} {\bibfnamefont {Baruch}\ \bibnamefont
  {Offir}}, \bibinfo {author} {\bibfnamefont {Yossi}\ \bibnamefont {Lev}}, \
  and\ \bibinfo {author} {\bibfnamefont {Rachel}\ \bibnamefont {Bezalel}},\
  }\bibfield  {title} {\enquote {\bibinfo {title} {Surface and deep learning
  processes in distance education: Synchronous versus asynchronous systems},}\
  }\href {\doibase 10.1016/j.compedu.2007.10.009} {\bibfield  {journal}
  {\bibinfo  {journal} {Computers \& Education}\ }\textbf {\bibinfo {volume}
  {51}},\ \bibinfo {pages} {1172--1183} (\bibinfo {year} {2008})}\BibitemShut
  {NoStop}%
\end{thebibliography}%

\end{document}